\documentclass[aps,twocolumn,preprintnumbers,amsmath,amssymb,floatfix,prb,superscriptaddress]{revtex4}

\usepackage{graphicx}
\usepackage{epstopdf}
\usepackage{dcolumn}
\usepackage{bm}
\usepackage[usenames]{color}
\usepackage{units}
\usepackage{float}
\usepackage{comment}

\graphicspath{{./}{./figs/}}

\begin{document}


\title{Topological Phase Transition Coupled with Spin-Valley Physics \\
in Ferroelectric Oxide Heterostructures}

\author{Kunihiko Yamauchi}
\affiliation{%
ISIR-SANKEN, Osaka University, 8-1 Mihogaoka, Ibaraki, Osaka, 567-0047, Japan }

\author{Paolo Barone}
\affiliation{%
Consiglio Nazionale delle Ricerche (CNR-SPIN), 67100 L'Aquila, Italy
}%
\affiliation{Graphene Labs, Istituto Italiano di Tecnologia, via Morego 30, 16163 Genova, Italy}

\author{Silvia Picozzi}
\affiliation{%
Consiglio Nazionale delle Ricerche (CNR-SPIN), 67100 L'Aquila, Italy
}%

\date{\today}
\newcommand{\ba}{Ba$_{2}$CoGe$_{2}$O$_{7}$}
\begin{abstract}
The possibility to engineer the coupling of spin and valley physics is explored in ferroelectric oxide heterostructures with $e_g^2$ electronic configuration. We show that the polar structural distortion induces the appearance of spin-valley coupled properties, at the same time being responsible for a topological transition from a quantum spin-Hall insulating phase to a trivial band insulator. The coupled spin-valley physics is affected by the topological band inversion in a non-trivial way; while the valley-dependent spin polarization of both conduction and valence bands is preserved, a change of the Berry curvature and of spin-valley selection rules is predicted, leading to different circular dichroic response as well as valley and spin Hall effects.


\end{abstract}


\maketitle


\section{Introduction}  

Spin-orbit coupling (SOC) is nowadays regarded as an important source for a rich variety of interesting and promising effects, laying, e.g., at the heart of topological quantum phases and spin-Hall effect\cite{TI_review2010,SHE_review2015}. In systems lacking inversion symmetry, SOC is generally responsible for spin-splitting Rashba and Dresselhaus effects at surfaces/interfaces and in the bulk, respectively\cite{Rashba1984,Dresselhaus1955};
recently, the existence of very large Rashba-like splitting in the bulk bandstructure has been reported in tellurohalides, a family of noncentrosymmetric pyroelectric semiconductors with strong SOC\cite{BiTeI, BiTeCl, Tellurohalides}.
On the other hand SOC may mediate spin-valley coupling in graphene-like hexagonal layered materials, which gives rise to valley-contrasting physics in the absence of inversion symmetry, as in the case of MoS$_2$ monolayer\cite{mos2_spinvalley,mos2_nature2014}, with appealing prospects in the field of spintronic and optoelectronic applications.

Since both spin-splitting and spin-valley effects may appear in acentric materials, it has been recently proposed to explore the properties of relativistic electronics in ferroelectrics, i.e., polar materials with switchable electric polarization\cite{silvia.review}. 
A large tunable Rashba effect has been predicted in distorted ferroelectric rock-salt chalcogenides GeTe and SnTe\cite{domenicoGeTe,Evgeny2014}, which can be considered as prototypes of a new class of multifunctional materials, where the permanent ferroelectric polarization could be used as a handle to control, in a non-volatile fashion, the Rashba-related properties such as the spin texture of the split bands. Recently, both the bulk giant Rashba splitting and the link between ferroelectricity and the spin polarization of the split bands of GeTe have been experimentally confirmed\cite{gete_exp1,gete_exp2}. 
On the other hand, the integration of ferroelectricity and spin-valley physics has proven to be more elusive. 

So far, the coupling of spin and valley degrees of freedom has been reported mostly in hexagonal layered or two-dimensional materials\cite{mos2_nature2014}. Even though semiconducting monolayers of group-IV elements, such as graphene, silicene and germanene, show an intrinsic spin-valley-sublattice coupling\cite{spinvalley.graphene,spinvalley.silicene, liu_2011}, the presence of inversion symmetry prevents the appearance of valley-contrasting effects; on the other hand, MoS$_2$ is a non-polar, albeit acentric, material, showing coupled spin-valley physics but no switchable electric polarization\cite{mos2_spinvalley}. Binary IV or III-V hexagonal monolayers have been predicted to develop a ferroelectric polarization when their structure is buckled, displaying both Rashba-like and Zeeman-like spin-split bands with coupled spin-valley physics analogous to MoS$_2$\cite{domenico_monolayer2015}. However, the spin-valley splitting was found to originate mainly from the presence of a diatomic basis in the honeycomb structure, being substantially unaffected by the reversal of the ferroelectricity, which only acts on the Rashba properties. On the other hand, the honeycomb structure can be engineered in heterostructures comprising bilayers of perovskite transition-metal oxides grown along the [111] direction\cite{nagaosa.natcom2011, okamoto_prl2013, okamoto_prb2014, hirai_2015}. This lattice geometry shows a number of potential advantages. First of all, it may give rise to Dirac points in the band structure of $d$ electrons (while most of the previous examples involve $s$ or $p$ electrons). Secondly, the symmetry of the crystalline field experienced by the $d$ electrons is reduced from octahedral -- causing a splitting in $e_g$ and $t_{2g}$ levels -- to trigonal, introducing additional level splittings of the transition-metal orbitals; such additional crystal-field effect may couple to SOC, trigger a spin-valley coupling and open the gap at the Dirac points, leading to quantum spin-Hall (or 2D $Z_2$ topological insulating) phases. Finally, ferroelectricity can be engineered in the heterostructure by sandwiching the bilayer in an insulating ferroelectric oxide; if the ferroelectric polarization is parallel to the growing direction [111], a non-volatile switchable layer potential difference can be in principle realized and easily manipulated, allowing to control and permanently tune the band-structure properties and, possibly, the spin-valley properties.

Based on these considerations, in a previous study we have proposed, by means of ab-inito materials design approach, to couple ferroelectricity and spin-valley properties in a BiAlO$_{3}$/BiIrO$_{3}$ perovskite heterostructure\cite{yamauchi.prl.biiro3}.
In this system, a BiIrO$_{3}$  bilayer -- realizing a buckled honeycomb lattice of Ir ions -- is embedded in the ferroelectric host BiAlO$_3$, a robust ferroelectric oxide which has been recently synthesized both as a thin film \cite{bialo3.son} and as a ceramic with a measured polarization $P_s\approx27~\mu$C/cm$^2$ along the perovskite [111] direction and a high critical temperature $T_{c}> 520^{\circ}$\cite{bialo3.zylbrberg}. The Ir low-spin 5$d^6$ manifold, comprising $t_{2g}$ states, is split by the trigonal crystal field in $a_{1g}$ and $e_g'$ states, analogously to the $\pi$ and $\sigma$ states of graphene; as a consequence, the low-energy properties of the valence band formed by $a_{1g}$ states in the paraelectric phase are accounted for by the same effective model proposed for buckled graphene (or silicene), underlying the analogies between the electronic properties of $p$ and $t_{2g}$ states. Additionally, the presence of a significant spin-valley-sublattice coupling gives rise, in the ferroelectric phase, to spin splitting and to a sizable net spin polarization $s_{z}$ at $K$ and $-K$ points in the hexagonal Brillouin zone that can be controlled by tuning the ferroelectric distortion. Consistenly with the early prediction of Xiao et al.\cite{nagaosa.natcom2011}, the band topology of this $t_{2g}^6$ system was found to be trivial.
%

In this contribution, we will take a step forward and design another bilayered oxide, exploring the possibility to engineer  the coupling of spin and valley physics also for $e_g$ electrons. Specifically, a $e_g^2$ system consisting of 
a LaAuO$_{3}$ bilayer embedded in (111) LaAlO$_{3}$ insulator
has been proposed to host a topological non-trivial band gap.
As shown in Sec. \ref{sec:model}, where we explicitly derive an effective model for $e_g$ electrons at valleys $K$ and $-K$, the topological insulating phase originates from a spin-valley-sublattice coupling that develops from the interplay of trigonal crystal field and an ``effective'' SOC arising from virtual excitations between $e_g$ and underlying $t_{2g}$ states; we further discuss how the ferroelectric polar distortions may trigger the appearance of coupled spin-valley physics,  analyzing their interplay with the topological properties. In this respect, a valley-polarized topological state has been recently proposed in a bismuth film.\cite{niu.bismuthfilm} On the other hand, spin-valley optical selection rules have been theoretically analysed in silicene when a perpendicular electric field is applied to the monolayer\cite{spinvalley.silicene}; the field was found to induce a topological transition from the quantum spin-Hall state to a trivial insulator, showing drastically different selection rules for photoexcited electrons. 

After discussing the expected low-energy electronic properties in Sec. \ref{sec:model} within a model Hamiltonian framework, in Sec. \ref{sec:abinitio} we adopt an ab-initio materials design approach to explore the interplay between structural distortions, topological properties and spin-valley physics in ferroelectric oxide heterostructures with $e_g^2$ electronic configuration.

\section{Effective model}\label{sec:model}

We start by discussing an effective low-energy model for $e_g$ electrons around point $K$ ($-K$) as derived from the tight-binding model previously introduced for the bilayered oxide\cite{nagaosa.natcom2011,yamauchi.prl.biiro3}. Hopping interactions between $e_g$ electrons are dominated by the strong $\sigma$ hybridizations $t^\sigma_{pd}$ of transition metal ions and oxygen ions, therefore the energy scale can be assumed to be $t_0=(t^\sigma_{pd})^2/\Delta_{pd}$, where $\Delta_{pd}$ is the level
difference between transition-metal $d$ orbitals and oxygen $p$ orbitals. 
In a trigonal setting with $z$ parallel to the polar axis [111], the symmetry adapted wavefunctions with original $e_g$ symmetry are defined as $\displaystyle{\vert L_1 \rangle=\sqrt{\nicefrac{2}{3}}\vert d_{yz}\rangle-\nicefrac{1}{\sqrt{3}}\vert d_{x^2-y^2}\rangle}$ and $\displaystyle{\vert L_2 \rangle=\sqrt{\nicefrac{2}{3}}\vert d_{zx}\rangle-\nicefrac{1}{\sqrt{3}}\vert d_{xy}\rangle}$ for each sublattice $L=A,B$. At valleys $K$ and $-K$ the $\sigma$-like hybridization between states $\vert L_i\rangle$ on different sublattices split the bands in a bonding/antibonding pair and in two non-bonding states. Each non-bonding state is related to a specific sublattice, being defined as:
\begin{eqnarray}
\vert\phi_1\rangle & = & \frac{1}{\sqrt{2}}\left(\vert A_1\rangle+i\,\tau\vert A_2\rangle \right)\nonumber\\
\vert\phi_2\rangle & = & \frac{1}{\sqrt{2}}\left(\vert B_1\rangle-i\,\tau\vert B_2\rangle \right),
\end{eqnarray}
where $\tau=\pm 1$ is a valley index labeling $\pm K$ points.
At odds with the previously analyzed $t_{2g}$ system, neither the SOC nor the trigonal crystal field directly affect $e_g$ electrons; however, they enter as a second-order effect via virtual excitations between $e_g$ and $t_{2g}$ levels.  In the symmetry-adapted basis, the effective SOC couples the $\vert L_1\rangle$ and $\vert L_2\rangle$ states with parallel spins, the corresponding matrix element being given by $i\tilde{\Lambda}\,s_z=i\,s_z\,2\Lambda^2(1/\Delta_{e_g'}-1/\Delta_{a_{1g}})$, where $\Lambda$ is the atomic SOC of the transition-metal ions, $s_z=\pm1$ denotes spin-up and spin-down components and $\Delta_{e_g'}=\Delta_o+\Delta_{t}$, $\Delta_{a_{1g}}=\Delta_o-2\Delta_t$ are the energy differences between $e_g$ and $e_{g}', a_{1g}$ levels, respectively, including the octahedral ($\Delta_o$) and trigonal ($\Delta_t$) crystal fields. Clearly, such effective SOC vanishes when $\Delta_t=0$ and all $t_{2g}$ levels are degenerate, underlying the importance of the trigonal crystalline field of the bilayered structure. However, the spin-up and spin-down components remain decoupled and $s_z$ is a good quantum number, implying that the non-bonding states are described by $\{\phi_1,\phi_2\}\otimes\{\uparrow,\downarrow\}$.
The dispersion around the valleys of these non-bonding states, including the effective SOC, reads as:
\begin{equation}\label{eq:h0}
H_0=\frac{\sqrt{3}}{2}t_0(\tau\,k_x\,S_x+k_y\,S_y)+\tilde{\Lambda}\,\tau\,s_z\,S_z,
\end{equation}
where $\bm s$ and $\bm S$ are the Pauli matrices describing the spin and the sublattice pseudospin, respectively.
While the first term clearly describes a Dirac point, the second term accounts for the spin-valley-sublattice coupling that is well known to open a gap in 2D Dirac semiconductors such as graphene or silicene, at the same time giving rise to hidden layer-dependent fully spin-polarized states at the valleys.  Due to the different parities of the half-filled molecular orbitals arising from the $e_g$ $\sigma-$hybridization at time-reversal-invariant moments $\Gamma$ and M, the $e_g^2$ system is a quantum spin-Hall insulator in the ideal bilayered perovskite structure.

A polar distortion along the [111] direction implies an offcentering $\delta z$ of the oxygens bridging the transition-metal ions perpendicular to the bilayer, thus opening new hybridization channels. While the change of the oxygen-mediated $e_g-e_g$ hopping is found to be an even function of the distortion $\delta z$, with the lowest-order correction being proportional to $(\delta z)^2$, new hopping interactions between the $e_g$ and $t_{2g}$ orbitals appear linear in $\delta z$. Keeping the lowest order contribution in $\delta z$ and $\Lambda$ in quasi-degenerate perturbation theory and neglecting $\pi$-like hybridization interactions, the effective hamiltonian describing the low-energy physics at valleys reads $H_0+\delta z\, H_1$, where
\begin{equation}\label{eq:h1}
H_1 = -E S_z + \alpha(\tau\,S_x\,s_y-S_y\,s_x)
\end{equation}
and $E=3\Lambda t_0^2/\Delta_{e_g'}^2$, $\alpha=2\Lambda t_0/\Delta_{e_g'}$.
The effect of a polar distortion is therefore analogous to that of an applied electric field perpendicular to a silicene monolayer. First, both the  two-fold degenerate valence and conduction bands at valleys are spin split, the top valence and bottom conduction bands showing substantially the same spin polarization which is opposite at time-reversal partners K and -K. Following Ezawa\cite{spinvalley.silicene}, the topological phase of the centric structure is expected to be lost when increasing the polar distortion (the electric field), which causes the valence band maximum and the conduction band minimum to be inverted at a critical distortion. The band inversion is expected to dramatically change the circular dichroic response due to different spin-valley optical selection rules, as well as to affect the Berry curvature describing the intrinsic contribution to the valley Hall effect when an electric field is applied in the bilayer plane. In the next Section we will explore these possibilities via an ab-initio approach in two perovskite oxide heterostructures.

\begin{figure}[th]
\begin{center}
{\center{
\includegraphics[width=86mm, angle=0]{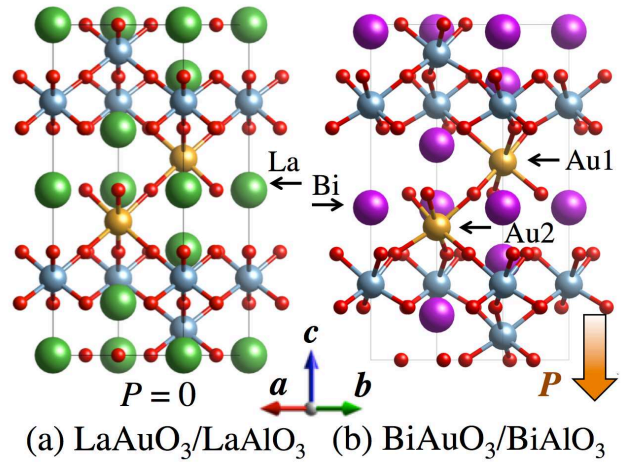}
\vspace {-0.6cm}
}}
\caption{\label{fig:crysband} 
Multilayer structure of (a) centrosymmetric  (LaAuO$_{3}$)$_{2}$(LaAlO$_{3}$)$_{n}$ and (b) polar (BiAuO$_{3}$)$_{2}$(BiAlO$_{3}$)$_{n}$. 
Bi ions are  simultaneously displaced with respect to the oxygen plane, causing the net polarization along the $c$ axis shown by a block arrow.  
}
\end{center}
\end{figure}
\begin{figure}[ht!]
\begin{center}
{
\includegraphics[width=86mm, angle=0]{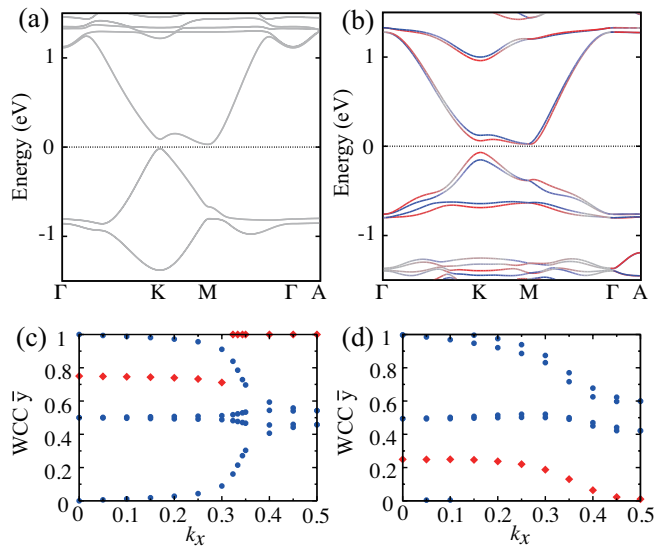}
}
\caption{\label{fig:bandwcc} 
Bandstructures for (a) centrosymmetric  (LaAuO$_{3}$)$_{2}$(LaAlO$_{3}$)$_{4}$ 
and (b) polar (BiAuO$_{3}$)$_{2}$(BiAlO$_{3}$)$_{4}$: $\pm s_{z}$ polarization projected on $E(k)$ curve is highlighted by red and blue color.
(c), (d) Evolution of hybrid Wannier charge centers (WCC) $\bar{y}$ versus $k_{x}$. 
The blue circles indicate the WCC of four occupied Au-$d$ states. 
The red diamonds indicate the middle of the largest gap between WCCs at given $k_{x}$. 
The resulted topological invariant $z_{2}$=1 and 0 for (LaAuO$_{3}$)$_{2}$(LaAlO$_{3}$)$_{4}$  and (BiAuO$_{3}$)$_{2}$(BiAlO$_{3}$)$_{4}$, respectively (see the main text). 
}
\end{center}
\end{figure} 

\section{Ab-initio results}\label{sec:abinitio}

Starting from the proposed topological heterostructure (LaAuO$_{3}$)$_{2}$(LaAlO$_{3}$)$_{n}$ and aiming at engineering ferroelectricity in the $e_g^2$ system,  we substitute La with Bi both in the insulating host and in the bilayer.
Density-functional-theory (DFT) calculations were performed using the VASP code \cite{vasp} with generalised gradient approximation potential. 
We optimized atomic structures  in a nonrelativistic scheme,  while ferroelectric polarization has been evaluated in the framework of the Berry-phase theory of polarization in 12 layers super cell, such as (BiAuO$_{3}$)$_{2}$(BiAlO$_{3}$)$_{10}$. 
Aiming at reducing the computational cost, we trimmed the less important host layers from the optimized supercell and take into account SOC in the six-layers supercell  (BiAuO$_{3}$)$_{2}$(BiAlO$_{3}$)$_{4}$. 
We checked that the cell reduction doesn't significantly modify the bandstructure of Au-$d$ states. 
Bandstructures and spin textures, including SOC, were plotted by using a 24$\times$24$\times$1 $k$-point mesh. 
In order to calculate the Berry curvature, we used Wannier90 code \cite{wannier90}, while
$Z_{2}$ invariants were calculated using the z2pack code\cite{z2pack}.

\subsection{Topological properties}

We first briefly summarize the structural properties of the considered multilayered systems shown in Fig. \ref{fig:crysband}. 
We found (LaAuO$_{3}$)$_{2}$(LaAlO$_{3}$)$_{10}$ and (BiAuO$_{3}$)$_{2}$(BiAlO$_{3}$)$_{10}$ to stabilize 
in the centrosymmetric $P\bar{3}m1$ and polar $R3c$ structure, respectively. 
(BiAuO$_{3}$)$_{2}$(BiAlO$_{3}$)$_{10}$ shows  a calculated polarization $P_{s}$ = 67.0 $\mu$C/cm$^{2}$.
As compared to the ideal perovskite 
structure, 
BiAlO$_{3}$ displays two main distortion modes, i.e., AlO$_{6}$ octahedral tilting and Bi-O polar distortion, the latter being responsible for the onset of ferroelectricity through the Bi lone-pair mechanism\cite{bifeo3.neaton}.
Au$^{3+}$-5$d^{8}$ electrons stabilize in a low-spin state in the O$_6$ octahedra, which are found to be sligthly compressed in the heterostructure geometry, thus realizing the anticipated trigonal crystal-field effect.
Bandstructures are shown in Fig. \ref{fig:bandwcc} (a) (b) for (LaAuO$_{3}$)$_{2}$(LaAlO$_{3}$)$_{4}$ and  (BiAuO$_{3}$)$_{2}$(BiAlO$_{3}$)$_{4}$, respectively. 
In both cases, an energy gap opens at the K points. 
In the centrosymmetric structure, the bands are spin-degenerated in all the $k$ points due to the inversion and the time-reversal symmetry, in agreement with Eq. (\ref{eq:h0}).
On the other hand, in the polar structure, the bands are spin split because of the additional coupling terms activated at the valleys $K$ by the structural distortions.
The topological $\mathbb{Z}_{2}$ number is evaluated by tracking the hybrid Wannier charge centers (WCC) \cite{vandelbilt.wcc1, z2pack}.
Figure \ref{fig:bandwcc} (c) (d) show the WCC evolution of four occupied Au-$d$ states in (LaAuO$_{3}$)$_{2}$(LaAlO$_{3}$)$_{4}$ and  (BiAuO$_{3}$)$_{2}$(BiAlO$_{3}$)$_{4}$, respectively. 
In the centrosymmetric case, the gap center (marked by red diamonds) jumps over one WCC at $k_{x}=$1/3, which corresponds to the $K$ point. 
It means that the band character (in this case its parity) is inverted at the Dirac cone, leading to a nontrivial topology. 
In the polar structure,  on the other hand, the largest gap makes no jumps and the band-structure topology is trivial.

\begin{figure}[ht!]
\begin{center}
{
\includegraphics[width=60mm, angle=0]{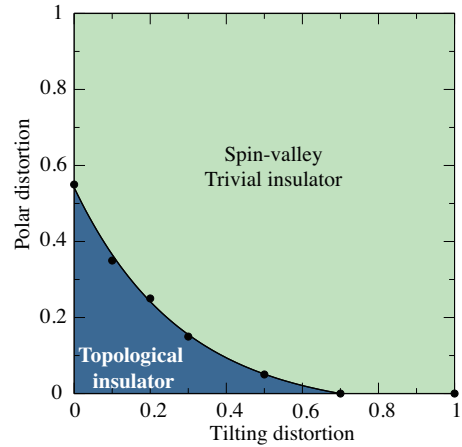}
}
\caption{\label{fig:phasediagram} 
Phase diagram of (LaAuO$_{3}$)$_{2}$(LaAlO$_{3}$)$_{10}$ under 
the polar and tilting distortion modes. 
The distortion amplitude comparing (BiAuO$_{3}$)$_{2}$(BiAlO$_{3}$)$_{10}$ and (LaAuO$_{3}$)$_{2}$(LaAlO$_{3}$)$_{10}$ is set as unit. 
  }
\end{center}
\end{figure}

\begin{figure}[ht!]
\begin{center}
{
\includegraphics[width=80mm, angle=0]{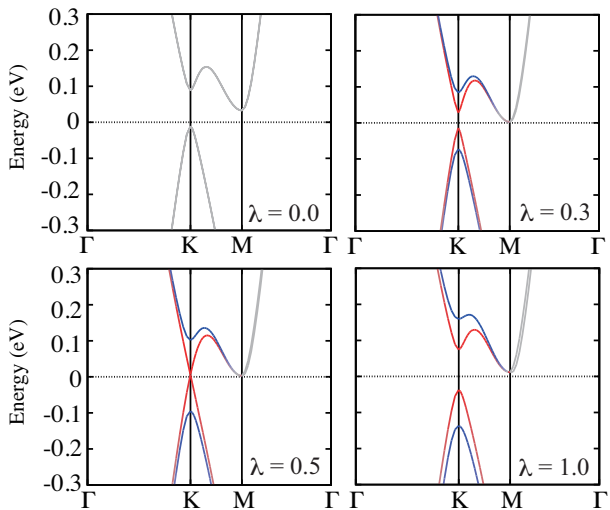}
}
\caption{\label{fig:banddevelop} 
Band-structure evolution of  (LaAuO$_{3}$)$_{2}$(LaAlO$_{3}$)$_{4}$ under a polar distortion parametrized by $\lambda$. 
The system is topological (trivial) insulator when $\lambda<0.5$ ($\lambda>0.5$). 
Red (blue) colour denotes up (down) $s_{z}$ component.  
  }
\end{center}
\end{figure}

In order to understand how the different structures affect the topological properties, we decomposed the structural distortion transforming the $P\bar{3}m1$ in the $R3c$ structure into symmetrical modes, tabulated in 
Table \ref{tab.isodistort}.
The largest distortion mode is $\Gamma_{1}^{-}$, which corresponds to the O$_{6}$ octahedron tilting mode, followed in order by  $\Gamma_{2}^{-}$, which is the polar distortion mode describing mainly Bi and O displacement. 
We then imposed the tilting and the polar distortion mode on (LaAuO$_{3}$)$_{2}$(LaAlO$_{3}$)$_{10}$ and evaluated the  $\mathbb{Z}_{2}$ invariants. 
As shown in Fig. \ref{fig:phasediagram}, 
both the tilting and polar modes causes a topological transition to a trivial insulating phase, the quantum spin-Hall phase being slightly more robust under the tilting than under the polar distortion. 
In Fig.\ref{fig:banddevelop} we show how the bandstructure is modified by the polar distortion.
The two-fold degenerate band at the K point is first spin-split, the $s_{z}$ component being linear to the polar distortion $\lambda$\cite{yamauchi.prl.biiro3}.
At $\lambda$=0.5, the spin-split bands with the same spin polarization at the valence and conduction bands touch, thus closing the energy gap. For larger distortions, $\lambda>0.5$, the gap opens again with inverted bands, making the topology of the bandstructure trivial, consistently with the effective model (cfr Eq. (\ref{eq:h1})).
Therefore only small polar distortions can allow the coexistence of topological properties and the spin-valley coupling. This situation may be realised by mixing La and Bi ions at A site; 
indeed, by considering a (Bi$_{1/6}$La$_{5/6}$)AuO$_3$ heterostrucure with La occupying five layers centered at the interface and Bi occupying other layers, the non-trivial topological phase is found to coexist with a weak polarization. 

\begin{table}[th]
\caption{
Structural distortion modes from (LaAuO$_{3}$)$_{2}$(LaAlO$_{3}$)$_{10}$ to (BiAuO$_{3}$)$_{2}$(BiAlO$_{3}$)$_{10}$.  
Irreducible representation of distortion mode, subgroup under the distortion, and the amplitude 
calculated by ISODISTORT.\cite{isodistort}  
}\label{tab.isodistort}
\vspace{0.2cm}
\hspace{-0.cm}\begin{tabular}{|l|c|c|c|}
\hline
 {\bf Mode} & {\bf Subgroup} & {\bf Ampl (\AA)} \\
\hline 
$\Gamma_{1}^{+}$ & $P\bar 3m1$ 	&0.20	\\
\hline
$\Gamma_{2}^{+}$ & $P\bar 3$		&0.10		\\
\hline
$\Gamma_{3}^{+}$ & $P\bar 1$		&0.00		\\
\hline
$\Gamma_{1}^{-}$ (tilting)  & $P321$		&1.57	  \\
\hline
$\Gamma_{2}^{-}$ (polar) & $P3m1$		&1.10		\\
\hline
$\Gamma_{3}^{-}$ & $P1$		&0.00		\\
\hline
\end{tabular}
\end{table}

%


\begin{figure}[ht!]
\begin{center}
\includegraphics[width=0.49\textwidth]{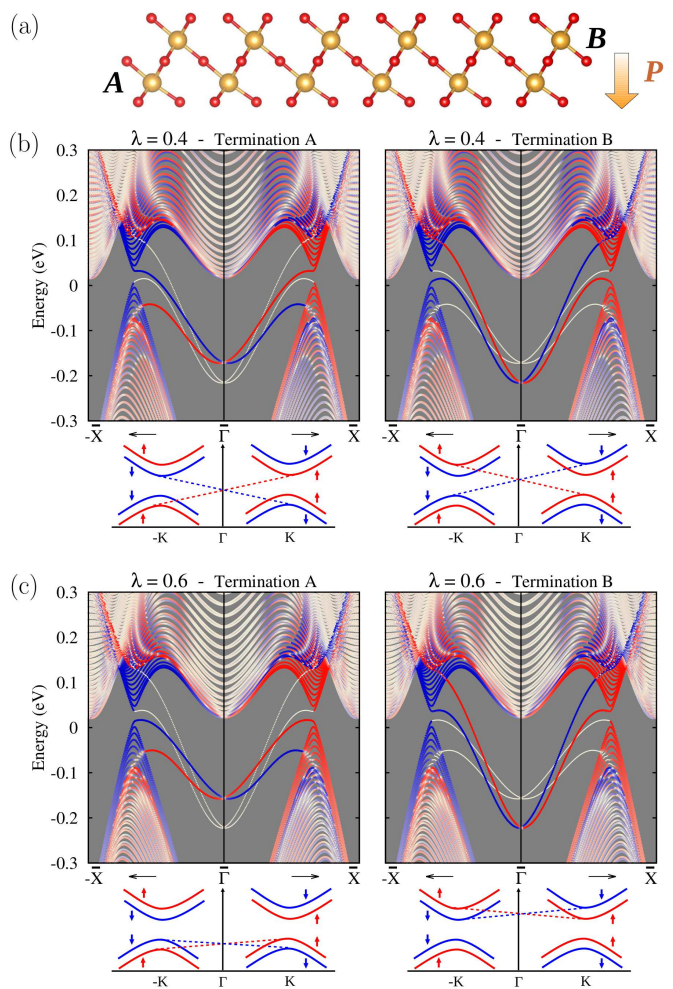}
\caption{\label{fig:edges} Edge states at inequivalent terminations -- sketched in (a) -- before (b) and after (c) the topological transition. Red (blue) color denotes up (down) $s_z$ component. Spin-polarised edge states always develop from spin-split bulk bands at opposite valleys, but cross the bulk gap only below the critical distortion $\lambda<0.5$, as schematically shown below each panel.}
\end{center}
\end{figure}

\begin{figure}[ht!]
\begin{center}
{
\includegraphics[width=85mm, angle=0]{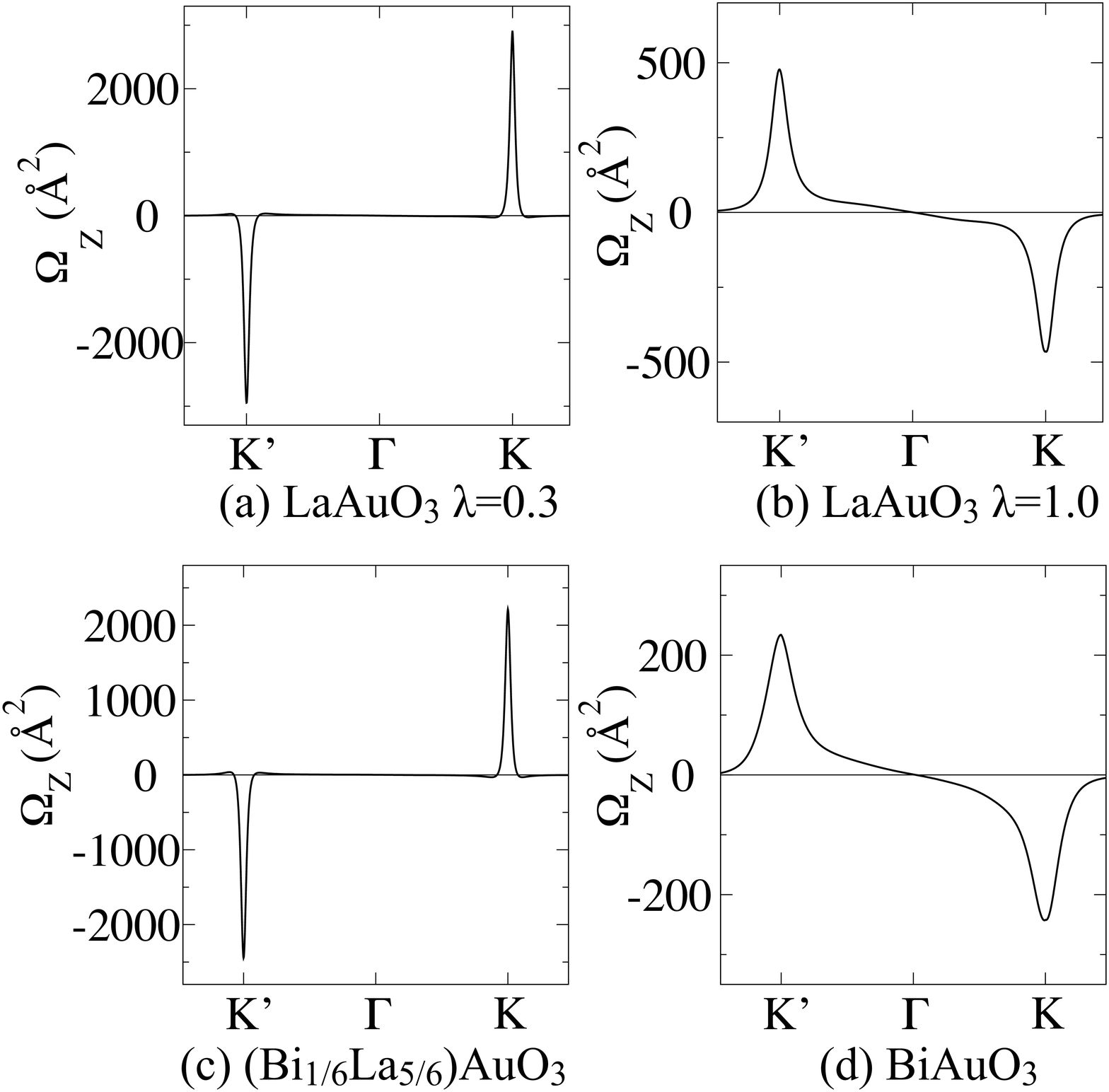}
}
\caption{\label{fig:berry} 
The $k_{z}$ component of Berry curvature $\Omega$ of occupied Au $e_{g}$ states in several bilayer oxide structures.  
(a) and (c) are topological while (b) and (d) are trivial insulators.
  }
\end{center}
\end{figure} 

We investigate then how the coupled spin-valley physics is modified at the topological transition. In Fig. \ref{fig:edges} we show the edge states of a zig-zag nanoribbon calculated in slab geometry from the real-space tight-binding model obtained by projecting the DFT bandstructure onto maximally localized Wannier functions\cite{wannier90}. The two terminations of the nanoribbon, consisting of Au ions belonging respectively to sublattice A or B, are inequivalent as soon as the polar distortion is activated, because of the lack of inversion symmetry. As a consequence, the edge states are strongly inequivalent at the two terminations, maintaining both the ionic character and the spin polarization of the corresponding sublattice. For instance, in the A-termination of the topological phase, a fully spin-polarized edge band developing from the bottom conduction bulk band around valley K connects to the second top valence bulk band at the time-reversed valley -K, with the same spin and sublattice character because of the spin-valley-sublattice coupling (see Fig. \ref{fig:edges}). For larger distortions $\lambda>0.5$ edge states are still found, which however, because of the bulk band-character inversion, do not cross the bulk energy band, connecting only valence (conduction) bands in the A (B) termination (see bottom panels of Fig. \ref{fig:edges}). In both cases, the strong bulk spin-valley coupling is reflected at the surface, causing the edge-band Bloch wavefunctions to be predominantly localized on a given sublattice depending on the spin and valley indices; therefore, spin-polarized edge states only develop from the corresponding spin-split bulk bands at the K (-K) point with the same $s_z$ component.

\subsection{Spin-valley physics at the topological transition}

The valley-dependent spin-polarization is not affected by the topological transition, while it is completely reversed when the ferroelectric polarization is switched. 
However, the bulk band inversion affects dramatically the Berry curvature $\Omega$, which gives rise to a Hall current under an applied in-plane electric field with a sign depending on the valley index\cite{mos2_nature2014}. We evaluated the dominant term of the Berry curvature using the Kubo-like expression\cite{vanderbilt.berry.2006}; 
\begin{equation}
\Omega_{\alpha\beta}=i\sum_{nm}(f_{m}-f_{n})\frac{H_{nm,\alpha}H_{mn,\beta}}{(\epsilon_{m}-\epsilon_{n})^{2}}. 
\end{equation} 
Peaks with opposite signs are found at time-reversed valleys, as shown in Fig. \ref{fig:berry}, i.e., the Berry curvature shows a strong valley contrast, implying that carriers at different valleys will move in opposite direction under an applied field, generating a valley Hall current. Due to the spin-valley coupling and the valley spin-polarization, such valley Hall effect would be accompanied by a spin-Hall effect.
When the band inversion occurs at the topological-trivial phase transition,  
the conduction and valence states exchanges the sign of the Berry curvature, thus leading to a reversal of the valley polarization but not of the spin polarization. As a consequence, the valley and spin Hall currents would be reversed at the topological transition, even though the spin-polarization of the carriers at the valleys is kept. 
The same situation is found when comparing topological Bi$_{1/6}$La$_{5/6}$AuO$_{3}$  and BiAuO$_{3}$ bilayer, where the peaks of $\Omega$ are broadened due to the AuO$_{6}$ octahedral tilting. 

\begin{figure}[ht!]
\begin{center}
\includegraphics[width=0.49\textwidth]{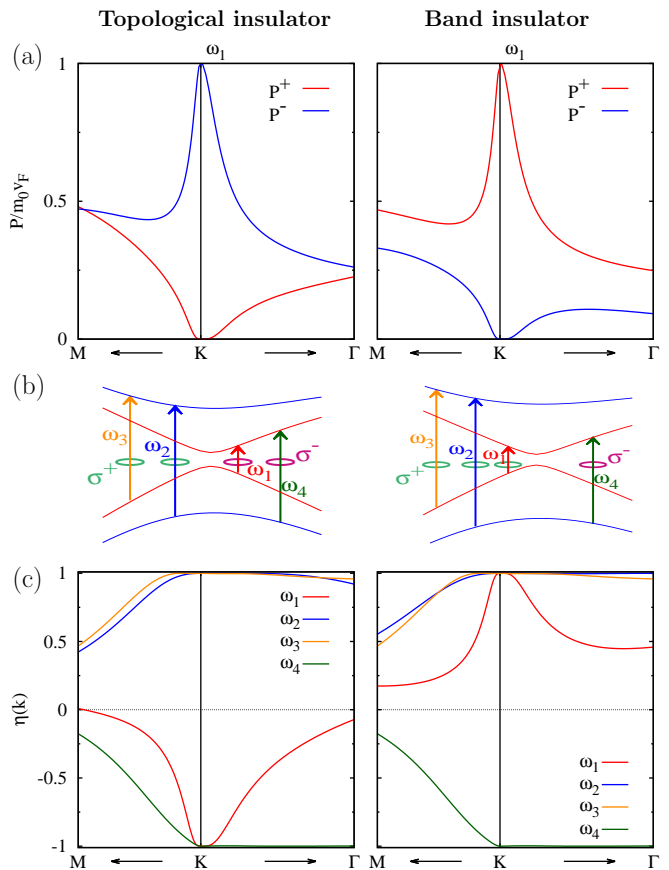}
\caption{\label{fig:optical} Optical selection rules for the topological($\lambda=0.4$, left panels) and band-insulating ($\lambda=0.6$ right panels) phases. The optical absorption for the fundamental transition $\omega_1$ under left- and right-handed circularly polarized light is displayed in (a). The optical circular polarization $\eta(k)$ for the interband transitions $\omega_i$ sketched in (b) as calculated in topological and band insulators is shown in (c) for a window of $k$-points around the valley $K$.}
\end{center}
\end{figure}

We also considered the valley-dependent selection rules for optical excitation with circularly polarized light. As depicted in Fig. \ref{fig:optical}(b), four interband transitions $\omega_i$ can be identified between two valence and two conduction states, $\omega_1$ labeling the fundamental transition between the top valence and bottom condution bands. The coupling strength with optical fields of  $\sigma^\pm$ circular polarization is given by $\mathcal{P}^\pm(\bm k)=\mathcal{P}^x(\bm k)\,\pm\,i\,\mathcal{P}^y(\bm k)$, where $\mathcal{P}^\beta(\bm k) = \nicefrac{m_0}{\hbar}\langle u_c(\bm k)\vert \nicefrac{\partial H}{\partial k_\beta}\vert u_v(\bm k)\rangle $ is the interband matrix element of the canonical momentum
operator and $m_0$ is the free electron mass. Due to time-reversal symmetry, the right-handed circular polarization at the K valley must be equal to the left-handed circular polarization at the -K valley, $\mathcal{P}^+(K)=\mathcal{P}^-(-K)$. 
As shown in Fig. \ref{fig:optical}(a), the dichroic response of the fundamental transition changes drastically at the topological transition,
leaving the other optical transitions qualitatively unchanged (not shown). In fact, the optical transition $\omega_1$ at valley $K$ is uniquely coupled with left-handed circularly polarized light in the topological phase, while it is coupled only with the right-circularly polarized optical field in the (trivial) band-insulating phase for $\lambda>0.5$, as displayed in  Fig. \ref{fig:optical}(a).

In order to further characterize the changes of the circular dichroism upon the topological transition, we evaluated the $k$-resolved optical polarization as the difference between the absorption of the right- and left-handed polarized lights normalized by the total absorption:
\begin{equation}\label{eq:optical}
\eta(k)=\frac{\vert\mathcal{P}^+(k)\vert^2-\vert\mathcal{P}^-(k)\vert^2}{\vert\mathcal{P}^+(k)\vert^2+\vert\mathcal{P}^-(k)\vert^2}.
\end{equation}
The calculated optical polarization $\eta(k)$ for all optical transitions around the valley $K$ point is shown in Fig. \ref{fig:optical}(c). All interband transitions are found to be perfectly polarized at the valley, implying that the optical selection rule holds exactly at $K$ points. It is worth to emphasize that only the optical polarization of the fundamental transition is reversed across the topological transition, being opposite whether the system is a topological or a band insulator.
As discussed in Ref.\cite{spinvalley.silicene}, this dramatic change of the optical response, occurring despite the valley-dependent spin polarization is unchanged, can be ascribed in fact to the inversion of the topmost valence band and the bottommost conduction band.



\section{Conclusions}
By combining an effective model analysis and DFT calculations, we explored the possibility to engineer and manipulate spin-valley coupling in the $e_g$ manifold of ferroelectric oxide heterostructures. Our study focused on Au-based oxide bilayers embedded in a ferroelectric insulating host, whose paraelectric counterpart with $e_g^2$ electronic configuration has been predicted to host a topological 2D insulating phase. Analogously to graphene and related 2D materials, the microscopic origin of this topological phase is ascribed to a spin-valley-sublattice coupling which arises from the interplay between the trigonal crystal field induced by the heterostructure geometry and an ``effective'' spin-orbit interaction 
due to virtual excitations between $e_g$ and $t_{2g}$ states of the transition-metal ions. We found that structural distortions, namely the polar and the O$_6$ octahedra tilting distortion modes, are detrimental for the quantum spin-Hall phase and cause a topological transition to a trivial band-insulating phase for moderately small distortions. On the other hand,  in the ferroelectric phase the spin-valley-sublattice coupling is responsible for the appearance of valley-contrasting phenomena, due to the breaking of inversion symmetry which allows to differentiate the time-reversed valleys $K$ and $-K$. Interestingly, the spin-valley properties are reflected in the symmetry-protected edge states in the quantum spin-Hall phase, at the same time being extremely sensitive to the topological transition induced by the polar distortion. Specifically, we found that the valley-contrasting Berry curvature - which describes the intrinsic contribution to the valley Hall effect - and the optical selection rule for the fundamental transition induced by circularly polarized light are opposite in the topological and band insulating phases. On the other hand, the spin polarization at the valleys is kept the same across the transition but is reversed when the ferroelectric polarization is switched. Hopefully, our theoretical predictions unveiling the non-trivial interplay between ferroelectricity, spin-valley physics and topological properties of $e_g$ electrons will stimulate further research in the field of functional oxide heterostructures.


\acknowledgments
KY acknowledges Bog G. Kim for his technical help to compute the hybrid WCC and the topological invariants. 
This work was supported by JSPS Kakenhi (No. 26800186 and 16H00993). PB acknowledges support from CNR Short-Term Mobility program Prot. AMMCNT-CNR 0054278 and the European Union's Horizon 2020 research and innovation programme under grant agreement No. 696656 GrapheneCore1.
The crystallographic figure was generated using VESTA program.\cite{vesta}
 
\bibliography{biblio}

\end{document}